
\magnification=1200\overfullrule=0pt\baselineskip=12pt
\vsize=22truecm \hsize=15truecm \overfullrule=0pt\pageno=0

\font\titlefont=cmbx12 scaled \magstep1
\font\sectnfont=cmbx10  scaled \magstep1
\font\subsectnfont=cmbx8  scaled \magstep1
\font\eightrm=cmr8
\long\def\fussnote#1#2{{\baselineskip=9pt
     \setbox\strutbox=\hbox{\vrule height 7pt depth 2pt width 0pt}%
     \eightrm
\footnote{#1}{#2}}}
\def\mname{\ifcase\month\or January \or February \or March \or April
           \or May \or June \or July \or August \or September
           \or October \or November \or December \fi}
\def\date{\hbox{\strut\mname \number\year}}
%
%
%
\newcount\FIGURENUMBER\FIGURENUMBER=0
\def\FIG#1{\expandafter\ifx\csname FG#1\endcsname\relax
               \global\advance\FIGURENUMBER by 1
               \expandafter\xdef\csname FG#1\endcsname
                              {\the\FIGURENUMBER}\fi}
\def\figtag#1{\expandafter\ifx\csname FG#1\endcsname\relax
               \global\advance\FIGURENUMBER by 1
               \expandafter\xdef\csname FG#1\endcsname
                              {\the\FIGURENUMBER}\fi
              \csname FG#1\endcsname\relax}
\def\fig#1{\expandafter\ifx\csname FG#1\endcsname\relax
               \global\advance\FIGURENUMBER by 1
               \expandafter\xdef\csname FG#1\endcsname
                      {\the\FIGURENUMBER}\fi
           Fig.~\csname FG#1\endcsname\relax}
\def\figand#1#2{\expandafter\ifx\csname FG#1\endcsname\relax
               \global\advance\FIGURENUMBER by 1
               \expandafter\xdef\csname FG#1\endcsname
                      {\the\FIGURENUMBER}\fi
           \expandafter\ifx\csname FG#2\endcsname\relax
               \global\advance\FIGURENUMBER by 1
               \expandafter\xdef\csname FG#2\endcsname
                      {\the\FIGURENUMBER}\fi
           figures \csname FG#1\endcsname\ and
                   \csname FG#2\endcsname\relax}
\def\figto#1#2{\expandafter\ifx\csname FG#1\endcsname\relax
               \global\advance\FIGURENUMBER by 1
               \expandafter\xdef\csname FG#1\endcsname
                      {\the\FIGURENUMBER}\fi
           \expandafter\ifx\csname FG#2\endcsname\relax
               \global\advance\FIGURENUMBER by 1
               \expandafter\xdef\csname FG#2\endcsname
                      {\the\FIGURENUMBER}\fi
           figures \csname FG#1\endcsname--\csname FG#2\endcsname\relax}
\newcount\TABLENUMBER\TABLENUMBER=0
\def\TABLE#1{\expandafter\ifx\csname TB#1\endcsname\relax
               \global\advance\TABLENUMBER by 1
               \expandafter\xdef\csname TB#1\endcsname
                          {\the\TABLENUMBER}\fi}
\def\tabletag#1{\expandafter\ifx\csname TB#1\endcsname\relax
               \global\advance\TABLENUMBER by 1
               \expandafter\xdef\csname TB#1\endcsname
                          {\the\TABLENUMBER}\fi
             \csname TB#1\endcsname\relax}
\def\table#1{\expandafter\ifx\csname TB#1\endcsname\relax
               \global\advance\TABLENUMBER by 1
               \expandafter\xdef\csname TB#1\endcsname{\the\TABLENUMBER}\fi
             Table \csname TB#1\endcsname\relax}
\def\tableand#1#2{\expandafter\ifx\csname TB#1\endcsname\relax
               \global\advance\TABLENUMBER by 1
               \expandafter\xdef\csname TB#1\endcsname{\the\TABLENUMBER}\fi
             \expandafter\ifx\csname TB#2\endcsname\relax
               \global\advance\TABLENUMBER by 1
               \expandafter\xdef\csname TB#2\endcsname{\the\TABLENUMBER}\fi
             Tables \csname TB#1\endcsname{} and
                    \csname TB#2\endcsname\relax}
\def\tableto#1#2{\expandafter\ifx\csname TB#1\endcsname\relax
               \global\advance\TABLENUMBER by 1
               \expandafter\xdef\csname TB#1\endcsname{\the\TABLENUMBER}\fi
             \expandafter\ifx\csname TB#2\endcsname\relax
               \global\advance\TABLENUMBER by 1
               \expandafter\xdef\csname TB#2\endcsname{\the\TABLENUMBER}\fi
            Tables \csname TB#1\endcsname--\csname TB#2\endcsname\relax}
\newcount\REFERENCENUMBER\REFERENCENUMBER=0
\def\REF#1{\expandafter\ifx\csname RF#1\endcsname\relax
               \global\advance\REFERENCENUMBER by 1
               \expandafter\xdef\csname RF#1\endcsname
                         {\the\REFERENCENUMBER}\fi}
\def\reftag#1{\expandafter\ifx\csname RF#1\endcsname\relax
               \global\advance\REFERENCENUMBER by 1
               \expandafter\xdef\csname RF#1\endcsname
                      {\the\REFERENCENUMBER}\fi
             \csname RF#1\endcsname\relax}
\def\ref#1{\expandafter\ifx\csname RF#1\endcsname\relax
               \global\advance\REFERENCENUMBER by 1
               \expandafter\xdef\csname RF#1\endcsname
                      {\the\REFERENCENUMBER}\fi
             [\csname RF#1\endcsname]\relax}
\def\refto#1#2{\expandafter\ifx\csname RF#1\endcsname\relax
               \global\advance\REFERENCENUMBER by 1
               \expandafter\xdef\csname RF#1\endcsname
                      {\the\REFERENCENUMBER}\fi
           \expandafter\ifx\csname RF#2\endcsname\relax
               \global\advance\REFERENCENUMBER by 1
               \expandafter\xdef\csname RF#2\endcsname
                      {\the\REFERENCENUMBER}\fi
             [\csname RF#1\endcsname--\csname RF#2\endcsname]\relax}
\def\refand#1#2{\expandafter\ifx\csname RF#1\endcsname\relax
               \global\advance\REFERENCENUMBER by 1
               \expandafter\xdef\csname RF#1\endcsname
                      {\the\REFERENCENUMBER}\fi
           \expandafter\ifx\csname RF#2\endcsname\relax
               \global\advance\REFERENCENUMBER by 1
               \expandafter\xdef\csname RF#2\endcsname
                      {\the\REFERENCENUMBER}\fi
            [\csname RF#1\endcsname,\csname RF#2\endcsname]\relax}
\def\refs#1#2{\expandafter\ifx\csname RF#1\endcsname\relax
               \global\advance\REFERENCENUMBER by 1
               \expandafter\xdef\csname RF#1\endcsname
                      {\the\REFERENCENUMBER}\fi
           \expandafter\ifx\csname RF#2\endcsname\relax
               \global\advance\REFERENCENUMBER by 1
               \expandafter\xdef\csname RF#2\endcsname
                      {\the\REFERENCENUMBER}\fi
            [\csname RF#1\endcsname,\csname RF#2\endcsname]\relax}
\def\Ref#1{\expandafter\ifx\csname RF#1\endcsname\relax
               \global\advance\REFERENCENUMBER by 1
               \expandafter\xdef\csname RF#1\endcsname
                      {\the\REFERENCENUMBER}\fi
             Ref.~\csname RF#1\endcsname\relax}
\def\Refto#1#2{\expandafter\ifx\csname RF#1\endcsname\relax
               \global\advance\REFERENCENUMBER by 1
               \expandafter\xdef\csname RF#1\endcsname
                      {\the\REFERENCENUMBER}\fi
           \expandafter\ifx\csname RF#2\endcsname\relax
               \global\advance\REFERENCENUMBER by 1
               \expandafter\xdef\csname RF#2\endcsname
                      {\the\REFERENCENUMBER}\fi
            Refs.~\csname RF#1\endcsname--\csname RF#2\endcsname]\relax}
\def\Refand#1#2{\expandafter\ifx\csname RF#1\endcsname\relax
               \global\advance\REFERENCENUMBER by 1
               \expandafter\xdef\csname RF#1\endcsname
                      {\the\REFERENCENUMBER}\fi
           \expandafter\ifx\csname RF#2\endcsname\relax
               \global\advance\REFERENCENUMBER by 1
               \expandafter\xdef\csname RF#2\endcsname
                      {\the\REFERENCENUMBER}\fi
        Refs.~\csname RF#1\endcsname{} and \csname RF#2\endcsname\relax}
\newcount\EQUATIONNUMBER\EQUATIONNUMBER=0
\def\EQ#1{\expandafter\ifx\csname EQ#1\endcsname\relax
               \global\advance\EQUATIONNUMBER by 1
               \expandafter\xdef\csname EQ#1\endcsname
                          {\the\EQUATIONNUMBER}\fi}
\def\eqtag#1{\expandafter\ifx\csname EQ#1\endcsname\relax
               \global\advance\EQUATIONNUMBER by 1
               \expandafter\xdef\csname EQ#1\endcsname
                      {\the\EQUATIONNUMBER}\fi
            \csname EQ#1\endcsname\relax}
\def\EQNO#1{\expandafter\ifx\csname EQ#1\endcsname\relax
               \global\advance\EQUATIONNUMBER by 1
               \expandafter\xdef\csname EQ#1\endcsname
                      {\the\EQUATIONNUMBER}\fi
            \eqno(\csname EQ#1\endcsname)\relax}
\def\EQNM#1{\expandafter\ifx\csname EQ#1\endcsname\relax
               \global\advance\EQUATIONNUMBER by 1
               \expandafter\xdef\csname EQ#1\endcsname
                      {\the\EQUATIONNUMBER}\fi
            (\csname EQ#1\endcsname)\relax}
\def\eq#1{\expandafter\ifx\csname EQ#1\endcsname\relax
               \global\advance\EQUATIONNUMBER by 1
               \expandafter\xdef\csname EQ#1\endcsname
                      {\the\EQUATIONNUMBER}\fi
          Eq.~(\csname EQ#1\endcsname)\relax}
\def\eqand#1#2{\expandafter\ifx\csname EQ#1\endcsname\relax
               \global\advance\EQUATIONNUMBER by 1
               \expandafter\xdef\csname EQ#1\endcsname
                        {\the\EQUATIONNUMBER}\fi
          \expandafter\ifx\csname EQ#2\endcsname\relax
               \global\advance\EQUATIONNUMBER by 1
               \expandafter\xdef\csname EQ#2\endcsname
                      {\the\EQUATIONNUMBER}\fi
         Eqs.~\csname EQ#1\endcsname{} and \csname EQ#2\endcsname\relax}
\def\eqto#1#2{\expandafter\ifx\csname EQ#1\endcsname\relax
               \global\advance\EQUATIONNUMBER by 1
               \expandafter\xdef\csname EQ#1\endcsname
                      {\the\EQUATIONNUMBER}\fi
          \expandafter\ifx\csname EQ#2\endcsname\relax
               \global\advance\EQUATIONNUMBER by 1
               \expandafter\xdef\csname EQ#2\endcsname
                      {\the\EQUATIONNUMBER}\fi
          Eqs.~\csname EQ#1\endcsname--\csname EQ#2\endcsname\relax}
%
\newcount\SECTIONNUMBER\SECTIONNUMBER=0
\newcount\SUBSECTIONNUMBER\SUBSECTIONNUMBER=0
\def\section#1{\global\advance\SECTIONNUMBER by 1\SUBSECTIONNUMBER=0
      \bigskip\goodbreak\line{{\sectnfont \the\SECTIONNUMBER.\ #1}\hfil}
      \bigskip}
\def\subsection#1{\global\advance\SUBSECTIONNUMBER by 1
      \bigskip\goodbreak\line{{\subsectnfont
         \the\SECTIONNUMBER.\the\SUBSECTIONNUMBER.\ #1}\hfil}
      \smallskip}
%
\def\lsim{\raise0.3ex\hbox{$<$\kern-0.75em\raise-1.1ex\hbox{$\sim$}}}
\def\gsim{\raise0.3ex\hbox{$>$\kern-0.75em\raise-1.1ex\hbox{$\sim$}}}
%

%
%
%
\def\binum{\hbox{BI-TP 91-06 \strut}}

\def\banner{\hfill\hbox{\vbox{\offinterlineskip
                              \binum\uctnum}}\relax}
\def\manner{\hbox{\vbox{\offinterlineskip
                        \uctnum\binum\date}}\hfill\relax}
\footline={\ifnum\pageno=0\manner\else\hfil\number\pageno\hfil\fi}
\def\binum{\hbox{BI-TP 92-45 \strut}}
\def\uctnum{\hbox{UCT-TP 187/92 \strut}}

\def\banner{\hfill\hbox{\vbox{\offinterlineskip
                              \binum\uctnum}}\relax}
\def\manner{\hbox{\vbox{\offinterlineskip
                        \uctnum\binum\date}}\hfill\relax}
\footline={\ifnum\pageno=0\manner\else\hfil\number\pageno\hfil\fi}
\footline={\ifnum\pageno=0\manner\else\hfil\number\pageno\hfil\fi}
\def\UCT{Department of Physics, University of Cape Town, Rondebosch 7700,
\par\noindent South Africa}
\def\BI{Fakult\"at f\"ur Physik, Universit\"at Bielefeld,
        W-4800 Bielefeld 1, Germany}
\def\Zagreb{Institute Ruder Bo\v{s}kovi\'c, Zagreb, Croatia}

{\vsize=21 truecm\banner\bigskip\baselineskip=10pt
\bigskip
\begingroup\titlefont\obeylines
\hfil {\bf Structure Functions of the Nucleon in a Statistical Model}
\hfil
\endgroup\bigskip
\bigskip
\centerline{J.~Cleymans$^{1}$,~
  I.~Dadi\'c$^{2,3}$,
  J.~Joubert$^1$ }
}
\bigskip\bigskip
\noindent {$^1$}\UCT\hfil\smallskip
\noindent {$^2$}\BI \hfil\smallskip
\noindent {$^3$}\Zagreb\hfil\smallskip
\bigskip
\bigskip
\bigskip
\centerline{\bf Abstract}
\medskip
\noindent Deep inelastic scattering is considered in
a statistical model of  the nucleon.
 This incorporates certain features which are absent in the
 standard parton model such as
 quantum
 statistical correlations which play  a role in the propagation
 of particles when considering Feynman diagrams containing internal lines.

 The inclusion of the ${\cal O}(\alpha_{s})$ corrections
in our numerical
 calculations allows a good fit to the data for
  $x\geq 0.25$.
 The fit corresponds to  values of temperature and
chemical potential of approximately
 $T=0.067$ GeV and $\mu=0.133$ GeV.
The latter values of parameters, however, give rise, for all $x$,
to a large value for $R=\sigma_{L}/\sigma_{T}$.
\vskip 1.5 truecm
\vfil\eject
\section{Introduction}
\vskip 6 pt
%
%
%
%
%
%
%
%
%
%
%
%
%
%
%
%
%
%
%
In this paper we further develop
a statistical model for the
 structure functions of the nucleon
[\reftag{ital},\reftag{g5},\reftag{aus},\reftag{indians}]
based on the MIT-bag model \ref{g6}.
The scattering of high energy leptons from a
 thermalized gas of quarks and gluons confined to the
volume of the nucleon was calculated by considering lowest-order diagrams
 in perturbative
 QCD. In refs.
[\reftag{zcd},\reftag{jcd},\reftag{ezawa},\reftag{cd}]
 these considerations were extended
 to the first order in the strong coupling constant $\alpha_{s}$
 since the effect of gluons is absent in zeroth order.

The phenomenological success of the parton model indicates that the number of
partons in the nucleon is very large due to
the $1/x$ behaviour
 of the structure functions at small $x$.

The parton model  neglects  quantum
statistical correlations due to the presence of
identical quarks and gluons in the initial and final states.
In the statistical model this is taken into account
through the use of Fermi-Dirac and Bose-Einstein distributions for
quarks and gluons, respectively. Stimulated emission factors for final-state
gluons and Pauli-blocking factors for final-state quarks are incorporated.
The propagation of particles through a many-body medium is taken into account
by using thermal Feynman rules for propagators and vertices.
These effects are negligible in lowest order, however, they are important
for higher order diagrams especially in the small $x$ region.

In ref. \ref{cd}, all processes
 contributing to order $\alpha_{s}$ (see Fig. 1)
to deep inelastic scattering of leptons off a heat bath of
quarks and gluons, were taken
into account exactly. All infrared, collinear
and ultraviolet divergences cancel in the framework provided by the
real-time formalism
of finite-temperature quantum field theory. The final
expressions from these analytical calculations are given in the
appendix.

The presence of the medium
leads to a dramatic increase in the length of the calculation.
For this reason many authors
[\reftag{bps},\reftag{aab},\reftag{ab},\reftag{aa},
\reftag{glm},\reftag{ggp}]
use special kinematics [e.g. in
lepton pair production from a thermalized quark gluon plasma
they would use a virtual photon with
zero spatial momentum: $q=(q_0,\vec{q}=0)$].
In our case we perform calculations for the case of general
kinematics [i.e. $q=(q_0,\vec{q}\neq 0)$].

 Results from analytical calculations of dilepton production from a
 quark-gluon plasma to order $\alpha_{s}$ ($q^2>0$ for the
 photon giving rise to the lepton
 pair), which is closely related to the process of deep inelastic scattering of
 leptons off a heat bath of quarks and gluons ($q^2<0$ for the photon stemming
 from the lepton beam), are presented in ref.~\ref{cd}.
 In Section~4 we present results
from numerical calculations of these expressions. There it is shown to what
extent recent deep inelastic scattering data can be reproduced by the
statistical model.
\vfil\eject
\section{Deep inelastic electron-proton scattering }
%
%
%
%
%
In our treatment of deep inelastic scattering
 we consider the scattering of a {\rm virtual}
 photon with 4-momentum $q=(\nu,0,0,q_z)$
 off the proton (considered as a heat bath of
 quarks and gluons).
 To order $\alpha_s$ the processes depicted in Fig. 1 contribute.
 According to Fig. 1
there can be more than two particles in the initial state
of a reaction so that the rate of reactions
(as opposed to a cross section) is the natural
 quantity to consider when relating the
measured structure functions on the hadronic level
with the results from our calculations of thermal averages of the fundamental
(virtual) photon-quark interaction processes.

The deep inelastic cross section for
 charged leptons can be written in such a form that
it depends linearly on $F_2$
 and non-linearly and very weakly on $R$ \ref{qbw}. Experimentally, $F_2$ and
$R$
 can be disentangled by measuring deep inelastic cross sections at the same
 $(x,Q^2)$ point but at different values of the incident beam energy. In
practice,
 as these cross sections depend only weakly on $R$, all measurements of $R$
 have large errors.

To each of the processes (a) to (f) in Fig. 1 there correspond
 the two Feynman diagrams in Fig. 2. These two diagrams,
 as drawn in Fig. 2 with exactly
 the same four-momentum assignments and directions of arrows, are used for
 each of the processes (a) to (f) in Fig. 1. When the energy
 components of the latter four-momenta
 are all positive the two Feynman diagrams refer to
 process (a) in Fig. 1 and the energy of a particle in diagrams
 (b) to (f) in Fig. 1 is considered to be negative when it
 changes from the left hand to the right hand side or vice versa relative
 to diagram (a) in Fig. 1.

 To each of the diagrams
 (g) and (h) in Fig. 1 there corresponds the six Feynman diagrams
 in Fig. 3. These six diagrams, as drawn in Fig. 3
 with exactly the same four-momentum assignments and directions of arrows, are
 used for each of the processes (g) and (h) in Fig. 1. For $k_0>0$
 and $k^{\prime}_0>0$ we are considering the three-particle process (g)
involving
 quarks and for $k_0<0$ and $k^{\prime}_0<0$ we are considering the
three-particle
 process (h) involving antiquarks.

The relation between the measured structure functions on the hadronic level
 and the results from our calculations of thermal averages of the fundamental
 (virtual) photon-quark interaction processes is discussed
 in refs. \ref{g5} and \ref{jcd}. One obtains on the hadronic level,
 as an example, for the ${\cal O}(\alpha_{s})$ process of gluon emission from a
 quark in Fig. 1(a):
$$
\eqalignno{
\sigma_{\lambda}^{(\rm{a})}
 =&{V\over 2K}\sum_q\int {d^3k\over (2\pi )^32|k_0|}n_F(x_k)
\int{d^3k'\over (2\pi )^32|k'_0|}
[1-n_F(x_{k'})]
&\cr
  &\times \int {d^3p\over (2\pi )^32|p_0|}[1+n_B(|p_0|)]
(2\pi )^4\delta^4(q+k-k^{\prime}-p)& \cr
 &\times {N_c^2-1\over 2}4\pi
\alpha_s4\pi\alpha e_q^2
\sum_{\rm{spins}}
|M_{\lambda}|^2       &(1)\cr
}
$$
according to the momentum assignments in Fig. 2
and where $V$ is the volume of
 the nucleon. In eq. (1) $\lambda$ refers to the helicity of the virtual
photon.
The matrix element for gluon emission from a quark
 for a given polarization, $\lambda$, of the space-like ($q^2<0$)
 photon ($\gamma^*$) is given by
$$
M_{\lambda}=M_1+M_2
\eqno(2)
$$
with
$$
\eqalignno{
M_1=&\bar{u}(k^{\prime},s^{\prime})
(-ig\gamma_{\mu})[\epsilon^{\mu}_{\rm{ g}}(p)]^*
iS^{11}(k+q)(ie\gamma_{\nu})\epsilon^{\nu}_{\lambda}(q)u(k,s) &(3)\cr
M_2
=&\bar{u}(k^{\prime},s^{\prime})(ie\gamma_{\mu})\epsilon^{\mu}_{\lambda}
iS^{11}(k-p)(-ig\gamma_{\nu})[\epsilon^{\nu}_{\rm{ g}}(p)]^*u(k,s)
&(4)\cr
}
$$
where we suppress reference to colour factors and fractional charges and where
$iS^{11}$ denotes the $11$-component of the fermion propagator in the real-time
formalism of finite-temperature quantum field theory.

The equations (3) and (4) provide explicit expressions for the matrix
element  in eq.~(1) (with
coupling constants,
colour factors and fractional charges factored out into explicit normalization
constants). We take $N_c=3$ and include up and down quarks in our model.

Fermion and boson distribution functions can be defined in
 such a manner that they are valid for both positive and negative energies.
They are for a fermion
$$ n_F(x_k)={1\over e^{\beta x_k}+1} \eqno(5) $$
where
$$ x_k=|k_0|-\mu\epsilon (k_0) \eqno(6) $$
where $\epsilon(k_0)$ gives the sign of $k_0$ and for a boson
$$ n_{\rm{ B}}(|k_0|)={1\over \rm{e}^{\beta |k_0|}-1}. \eqno(7) $$
\section{Calculation of the structure functions}
The zeroth order in $\alpha_s$ contributions can be calculated in a
straightforward
manner \ref{g5}. The calculation of ${\cal O}(\alpha_s)$
 contributions require a much larger effort. In the
 calculations of the ${\cal O}(\alpha_{\rm{ S}})$ contributions, the
 phase space and loop integrations can be
 analytically reduced \ref{cd} to an integration over the two energy variables
$K$
 and $p_0$ as shown in the final expressions for four-particle processes
 given by eqs. (B1) and (B2) and for
 three-particle processes given by eqs. (C1)
and (C2). These
 double integrals cannot be analytically calculated. Their numerical
 calculation is discussed in Section~4.2.

We distinguish between the zeroth and first order in $\alpha_s$
 contributions by writing:
$$
F_2=F^{(0)}_2+F^{(1)}_2. \eqno(8)
$$
Since we are also interested in the structure function
$R$ which is expressible
 in terms of $F_2$ and $F_1$ , we also calculate
$$
F_1=F^{(0)}_1+F^{(1)}_1.
\eqno(9)
 $$
Zeroth order in $\alpha_s$ expressions for $F_2$ and $F_1$
 in the statistical model are given in ref. \ref{g5}:
$$
\eqalignno{
F^{(0)}_2
=&  {3M^2x^2V\over 2\pi^2}
{\left(1+Mx/2\right)^2\over \left(1+Mx/\nu\right)^5}
\sum_q e_q^2 \left\{ \left[ {3\over 2}
\left(1+{2\mu_q\over \nu}\right)^2
-{1\over 2}\left(1+{Mx\over \nu}\right)^2\right]
f_0(z_q)\right.        &     \cr
&\left.+{6T\over \nu}\left(1+{2\mu_q\over \nu}\right)f_1(z_q)+
{6T^2\over \nu^2}f_2(z_q)+\mu_q\rightarrow
-\mu_q,z_q\rightarrow \bar{z}_q\right\}
&(10)\cr
F^{(0)}_1=&{3M^2x^2VT\over 8\pi^2}{\left(1+Mx/2\right)\over
\left(1+Mx/\nu\right)^3}\sum_qe_q^2\left\{\left[\left(1+
{2\mu_q\over \nu}\right)^2+\left(1+{Mx\over \nu}\right)^2\right]
f_0(z_q)\right.  &\cr
&\left.+ {4T\over \nu}\left(1+{2\mu_q\over \nu}\right)
f_1(z_q)+{4T^2\over \nu^2}
f_2(z_q)+\mu_q\rightarrow -\mu_q,z_q\rightarrow \bar{z}_q\right\}
&(11)\cr
}
$$
where the sum over $q$ is over quark flavours only and where
$$
z_q\equiv {Mx\over 2T}- {\mu_q\over T}
\eqno(12)
$$
and
$$
\bar{z}_q\equiv {Mx\over 2T}+{\mu_q\over T}
\eqno(13)
$$
$$
f_n(z)\equiv \int_z^{\infty}{y^ndy\over e^y+1} .
\eqno(14)
$$
The ${\cal O}(\alpha_s)$ parts of expressions for $F_2$ and $F_1$
 in the statistical model are:
$$\eqalignno{
F^{(1)}_2
=&{K\over 4\pi^2\alpha}(\sigma^{(\alpha_{\rm{ S}})}_{\rm{ T}}
   +\sigma^{(\alpha_{\rm{ S}})}_{\rm{ L}})
{-q^2\nu\over q_z^2}
&(15)\cr
F^{(1)}_1=&{K\over 4\pi^2\alpha}M\sigma^{(\alpha_{\rm{ S}})}_{\rm{ T}}
&(16)\cr
}$$
where
$$
\eqalignno{
\sigma^{\alpha_s}_T
&={1\over 2}{V\over 2K}{N_c^2-1\over 2}
4\pi\alpha_s 4\pi\alpha \sum_q e_q^2\left\{
\int \bar{M}_{\Sigma}S_4d\mu
+\sum_{W=B,F,F'}
\int \bar{M}_{\Sigma}^{3 W}S_{3 W}
d\mu_W \right.&\cr
&\left.
+\int \bar{M}_0S_4d\mu +\sum_{W=B,F,F'}\int \bar{M}_0^{3W}S_{3W}d\mu_W
 \right\}&(17)\cr
\sigma^{\alpha_s}_L
=&{V\over 2K}
{N_c^2-1\over 2}4\pi\alpha_s4\pi\alpha
\sum_q e_q^2\left\{\int
\bar{M}_0S_4d\mu
+\sum_{W=B,F,F'}\int \bar{M}_0
^{3W}S_{3W}d\mu_W    \right\}
&(18)\cr
}$$
with the expressions in curly brackets given in
eqs.~(C1), (C2), (B1) and (B2). The have to be
 evaluated for each flavour of quark, since the chemical
 potential appearing in the statistical factors $S_4$ and $S_{3W}$
 differ for each flavour as we sum over the flavours of quarks.
 As explained in Appendices~B and ~C, the $K,p_0$ integration regions for
 four- and three-particle processes are as given in Figs. 4 and 5,
 respectively,
and the integrand for four-particle processes is different in each
 subregion shown in Fig. 4.
\section{Numerical Evaluation}
The numerical calculation of the
 structure functions $F_2$ and $R$ was performed in two parts.
 The  zeroth order is discussed in the next section while the
contributions of
 ${\cal O}(\alpha_s)$  are discussed in
 Section~4.2. Results from the total
 calculation, i.e., zeroth order plus ${\cal O}(\alpha_s)$
 results are presented in Section~4.3.
\vskip 3pt
\subsection{Zeroth Order}

We chose to fit to a plot of $F_2$ versus $x$,
which is determined by
 parton distributions which reproduce a wide range of experimental data.
 Such parton distributions were published in 1990 by Kwiecinski et al.
 in ref. \ref{we}. They parametrize the
 distributions at $Q^2_0\equiv -q^2=4\rm{\ GeV}^2$ and evolve up in $Q^2$
 in order to test their parton distributions against experimental data.
 We fit to their so-called $B_{\_}$ fit at $Q^2=4\rm{\ GeV}^2$.

 The values for temperature and chemical potential which give the best
 fit for $F_2$ versus $x$ are used to calculate the corresponding plot of $R$
 versus $x$ which we will compare with
 experimental data given by Whitlow \ref{wh}.
%
%
%
%
%
 In the deep-inelastic scattering
 limit, $x,T,\mu_q$ finite with $\nu$ going to infinity, the expression for
 $F_2^{(0)}$ in eq. (10) is dominated by the $f_0$ term which can be
 evaluated analytically.
In Ref. \ref{g5} the latter expression
was used to fit  the experimental data
by Aubert et al. \ref{qbl}.

In this work we keep the full
 expression for $F^{(0)}_2$ as given in eq. (10) and fit to
 $F_2$ as determined by the parton distributions
by Kwiecinski et al. in ref. \ref{we} (their $B_-$ fit).
The values for temperature and chemical potential
which produce the best fit to the plot of $F_2$ versus $x$ are
used to calculate $F_1^{(0)}$ in eq. (11)
 too.
Then the structure function
$R$ which is expressible in terms of $F_2^{(0)}$
 and $F_1^{(0)}$ can be calculated.
%
%
%
%
%
%
%
%
 The chemical potential for the down quarks was kept fixed in terms of the
 chemical potential for the up quarks according to
$$
\mu_{\rm{ down}}={\mu_{\rm{ up}}\over 2^{1/3}}\eqno(19)
$$
(as was done in ref.~\ref{g5}) in order to reproduce the ratio of up to
 down quarks as being 2 to 1. This is approximately the case for a quark
 gas at very low temperatures.

As opposed to the $f_0$ function, the functions $f_1$ and $f_2$, which also
 appear in the expression for $F_2^{(0)}$ in eq. (10),
cannot be evaluated
 analytically. They were
evaluated numerically.

For $x\geq 0.4$, the zeroth order in $\alpha_s$ calculation produced
 a good fit to the $Q^2=4 \rm{ GeV}^2$ data of Kwiecinski et al. as shown in
 Fig. 6.

According to the following rough estimates, the failure of the zeroth
 order in $\alpha_s$ theory of the statistical model to reproduce
 the low $x$ behaviour of $F_2$ could be
 due to the finite volume of the nucleon.
{}From the mass shell condition on the four-momentum $k'=k+q$ of the outgoing
 quark in the zeroth order in $\alpha_s$
(lowest order Feynman diagram in Fig. 3),
one obtains a lower limit on $|\vec{k}|$ of the incoming
 quark as follows. From the mass shell condition for {\rm massless} quarks
 [$(k+q)^2=0$] and the relation $x=Q^2/(2M\nu)$ one obtains
$$
-1\ \ \ \leq\ \ \ \cos\theta_{\vec{k}\vec{q}}={2|\vec{k}|\nu-Q^2
\over 2|\vec{k}|
\sqrt{\nu^2+2M\nu x}}\ \ \ \leq\ \ \  1.
\eqno(20)
$$
The first inequality in the latter line produces
$$
\eqalignno{
|\vec{k}|\geq &{Q^2\over 2\nu +2\nu\sqrt{1+{2Mx\over \nu}}}=
{2M\nu x\over 2\nu +2\nu\left[1+{\cal O}\left(
{x\over \nu}\right)\right]}&\cr
=&{Mx\over 2}+{\cal O}\left({x\over \nu}\right).
&(21)\cr
}
$$
Another lower limit on $|\vec{k}|$ is obtainable from
the Heisenberg uncertainty
 principle of the general
form $\Delta x\Delta k\geq 1$. Taking $\Delta x$ equal
 to the diameter of the nucleon (2 fermi = 10.14 $\rm{GeV}^{-1}$),
one obtains
 $\Delta k\geq (10.14)^{-1}$. From this and the value of the lower limit on
 $|\vec{k}|$ in eq. (21) one could deduce that our model is only valid
 for $x$ values satisfying
$$
{Mx\over 2}\geq (10.14)^{-1},
\eqno(22)
$$
i.e., $x\geq 0.21.$

The zeroth order in $\alpha_s$ results for the structure function
$R$ are shown in Fig. 7. For $x\geq 0.25$ these results are consistent
with the $Q^2=4 \rm{ GeV}^2$ experimental data given in Ref. \ref{wh}.

The zeroth order in $\alpha_s$ calculation is only able
 to reproduce the data for values of $x\geq 0.4$. In
 Section~4.2 we investigate to what extent the ${\cal O}(\alpha_s)$
 corrections are able to extend the fit
 to values of $x$ smaller than 0.4.

\subsection{The zeroth- plus first order in $\alpha_s$
 calculation }
In this section we take the ${\cal O}(\alpha_s)$ corrections into
 account by numerically calculating the quantities given in eqs.
(15) and
 (16) and adding them to zeroth order in $\alpha_s$ results
 according to eqs. (8) and (9). The zeroth order in
 $\alpha_s$ quantities given in eqs. (10) and (11) are
 calculated in the manner discussed in Section~4.3. Due to the
 modifications introduced by the ${\cal O}(\alpha_s)$ contributions,
 we search for new values of temperature and chemical potential which will
 produce a fit to the plot of $F_2$ versus $x$ of Kwiecinski et al. given in
 Fig. 6$(a)$.

 We keep the
 chemical potential of the down quarks fixed
in terms of the  chemical potential
 of the up quarks according to eq. (19). The values for temperature and
 chemical potential which produce the best fit to the data
are used to calculate $F_1$ in eq. (9).
 Then the structure function $R$ which is expressible in terms of $F_2$
 and $F_1$ can be calculated and compared
 to the experimental data.
%
%
%
%
%
%
%
%

%
According to the ${\cal O}(\alpha_s)$ expressions, $F_2^{(1)}$
 and $F_1^{(1)}$ in eqs. (15) and (16),
one needs to numerically
 calculate the double integrals in eqs. (B1), (B2),(C1) and
 (C2).
The $K,p_0$ integration regions for four- and three-particle
 processes are shown in
Figs. 4 and 5, respectively. These
 integrals are devoid of ultraviolet, infrared and collinear singularities
 as discussed in ref.~\ref{cd}.
 However, individual terms in the integrands can still become infinite along
 certain lines in the $K,p_0$ integration regions. This happens when factors
 in denominators or arguments of $\ln$ functions in some terms become equal
 to zero along certain lines.

Examples are singularities which arise along the line along which
 the quantity $b$ is equal to zero. The latter singularity arises in the terms
 proportional to $\ln |\bar{C}_4|$ and $|G_4|$ in eq. (B1).
One of the subregions which contain the line along which $b$ is equal to zero
 is subregion D in Fig. 4.

Even though singularities from such terms cancel each other,
the terms individually give rise to singularities.
Therefore
 one should take care not to approach the line along which $b=0$ too closely
 in the numerical integration. For the same reason one should steer clear
 of other lines along which other individual terms in expressions for
 three- and four-particel processes are singular. This is accomplished by
 implementing a trapesium integration method whereby the integrand is
 evaluated at the discrete points of
  a two-dimensional lattice constructed in such a way that the two
 nearest successive lattice points to any of the lines along
 which singularities arise, are always located an equal distance on both
 sides of the relevant line. The value of the integral is the stable
 value attained in
 successive numerical integrations in which, independently and in steps, the
 density of lattice points is increased and the range of integration is
 extended (The presence of thermal distribution factors in terms suppresses
 contributions at large values of the integration variables.).
\subsection{Results}
In Section~4.1 it was mentioned that the zeroth order in
 $\alpha_s$ calculation is able
 to reproduce the data of Kwiecinski et al. on $F_2$ versus $x$ for values of
 $x\geq 0.4$ only. From Fig. 8 it can be seen that, by including
 ${\cal O}(\alpha_s)$ corrections, the reproduction of the data of
 Kwiecinski et al. could be extended to values of $x$ smaller than 0.4. A
 reasonable fit could be obtained for values of $x\geq 0.25$. This is
compatible
 with our estimate for the limited range of validity of our theory
 ($x\geq 0.21$ as derived in Section~4.1) due to
the finite volume of the nucleon.

For the fixed values of the parameters and the fixed expression
 for the chemical potential of the down quarks in terms of the chemical
potential
 of the up quarks in eq.~(19), a fit to the data of Kwiecinki et al. is
 possible for values of temperature and chemical potential in the immediate
 vicinity of $T=0.067$ GeV and $\mu_{\rm{ up}}=0.133$ GeV only.

Figs. 9 and 10 show how the results
 for $F_2$ versus $x$ from the calculation including zeroth order plus
 ${\cal O}(\alpha_s)$ contributions vary as a function of temperature
 at fixed chemical potential and as a function of chemical potential
 at fixed temperature, respectively.

{}From Fig. 8 it can be seen that the ${\cal O}(\alpha_s)$
corrections to $F_2$ is negative for large $x$. Such negative values could in
principle arise from interference terms from three-particle processes since,
by neglecting the ${\cal O}(\alpha^2_s)$ contributions
to the amplitude squared of the
${\cal O}(\alpha^0_s)$ and
${\cal O}(\alpha^1_s)$ Feynman diagrams for three-particle
processes (see Fig. 3),
the positive definiteness of the amplitude squared is lo

Results of the calculation of the structure function $R$ with
 zeroth order plus ${\cal O}(\alpha_s)$ expressions for $F_2$ and
 $F_1$, as they appear in the expression for $R$ are shown in
 Fig. 11. In the $x\geq 0.25$ region of interest the results
 typically lie a factor of 6 above the experimental
 data for $R$. The absence of data for $x\geq 0.8$
 in Fig. 11 is due to the numerical instability encountered as $F_1$,
  appearing in a denominator in the expression for $R$,
goes to zero
 at large $x$.
\section{Summary}
In this work, we considered deep inelastic scattering of leptons off a proton
in
 the statistical model  proposed in ref. \ref{g5}. The interior of
 the nucleon is viewed as a
thermalized assembly of up and down quarks and gluons.
 This enables one to incorporate features which are absent in the standard
 parton model. Quantum statistical correlations are incorporated
 through the use of Fermi-Dirac
and Bose-Einstein distributions for initial-state
 quarks and gluons, respectively. Stimulated emission factors for final-state
 gluons and Pauli-blocking factors for final-state quarks are incorporated.
 The propagation of particles through a many-body medium is taken into account
 by using thermal Feynman rules for propagators and vertices. The statistical
model
 could also be seen as an attempt to describe the interior of
 the nucleon at a more fundamental level than that attained through the use
 of arbitrary parton distributions containing many
parameters in the parton model.

We chose to fit to a plot of $F_2$ versus $x$,
which is determined by
 parton distributions which reproduce a wide range of experimental data.
 Such parton distributions were published in 1990 by Kwiecinski et al.
 in \ref{we}. They decide on a parametrization of the parton
 distributions at $Q^2_0\equiv -q^2=4 \rm{\ GeV}^2$ and evolve up in $Q^2$
 in order to test their parton distributions against experimental data.

Our zeroth order in the strong coupling constant $\alpha_{\rm{ S}}$
 calculation of the structure function $F_2$ versus $x$ at $Q^2=4\rm{ GeV}^2$
produced a
 good fit to the data  for $x\geq 0.4$. The values for
 temperature and chemical potential which give the best
 fit are $T=0.04$ GeV and $\mu_{\rm{ up}}=0.21$ GeV (with
 $\mu_{\rm{ down}}=\mu_{\rm{ up}}/2^{1\over 3}$).
 Other combinations of values for $T$ and $\mu_{\rm{ up}}$ do not produce
 an improved fit to the data. In consequence, we investigate
 to what extent ${\cal O}(\alpha_{\rm{ S}})$
 corrections are able to extend the reproduction of the data
 to values of $x$ smaller than 0.4. The corresponding zeroth order in
 $\alpha_{\rm{ S}}$ results for the structure function $R$ are
 consistent with experimental data Fig. 7.

The inclusion of the ${\cal O}(\alpha_s)$ corrections \ref{jcd}
 in our numerical calculations allowed the extension of the
 reproduction of the data of Kwiecinski et al. to values of $x$ smaller than
0.4. A
 fit could be obtained for values of $x\geq 0.25$. This is compatible
 with our estimate for the limited range of validity of our theory
 ($x\geq 0.21$) due to a finite-size effect arising from the finite volume of
the nucleon.
 The latter fit is
 possible for values of temperature and chemical potential in the immediate
 vicinity of $T=0.067$ GeV and $\mu_{\rm{ up}}=0.133$ GeV only (with
 $\mu_{\rm{ down}}=\mu_{\rm{ up}}/2^{1\over 3}$
and $\alpha_s=0.2$).

The latter values of parameters, however, do not fit  the
 structure function $R=\sigma_{\rm{ L}}/\sigma_{\rm{ T}}$ to the
 experimental data presented in Ref. \ref{wh}.
Even when taking into account
 that all measurements of $R$ suffer from large experimental errors due to
 the weak dependence of the deep inelastic cross section for charged
 leptons on $R$, the size of the
discrepancy remains unacceptable. This indicates
 a shortcoming of the statistical model in its present form to reproduce
 the  structure function of the proton.
\bigskip
\bigskip
\bigskip\bigskip
{\bf Acknowledgment}
\medskip
We acknowledge financial support from the KFA (Juelich), the European
Economic Community under contract No. CI1$^*$-CT91-0893 (HSMU)
and from the Foundation for Research Development (Pretoria).  We gratefully
acknowledge the hospitality of the Physics Department of the University
of Bielefeld where this work was finalized.
\vfil\eject
\bigskip\noindent
{\bf Appendix A : Final expressions for three- and four-particle processes}
\medskip
In this appendix we summarize expressions derived for the phase space
 integrations
 of the matrix element squared from four- and three-particle processes.

The way in which the final expressions for four-particle processes
in eqs. (B1)
 and (B2) and for three-particle processes in eqs. (C1)
 and (C2) contribute to the structure
 functions of the nucleon can be seen by considering eqs. (8),(9)
 and (15)--(18).

In order to write the expressions in a simplified form, we introduce
the abbreviations
$$
\eqalign{
A=&(2K+p_0+\nu )/2                            \cr
B=&(2K-p_0+\nu )/2                            \cr
C=&(2K+p_0+q_z\epsilon (k_0k_0^{\prime}p_0))/2\cr
D=&(2K-p_0+q_z\epsilon (k_0k_0^{\prime}p_0))/2\cr
E=&(\nu +q_z\epsilon (k_0k_0^{\prime}p_0))/2  \cr
}
\qquad
\eqalign{
a=&(2K+p_0-\nu )/2                                  \cr
b=&(2K-p_0-\nu )/2                                  \cr
c=&(2K+p_0-q_z\epsilon (k_0k_0^{\prime}p_0))/2\     \cr
d=&(2K-p_0-q_z\epsilon (k_0k_0^{\prime}p_0))/2      \cr
e=&(\nu -q_z\epsilon (k_0k_0^{\prime}p_0))/2\quad (A1) \cr
}
$$

where $\epsilon$ is the sign function.

 In the calculation of the ${\cal O}(\alpha_s)$
 corrections \ref{cd}, all phase space integrations are analytically
 reduced to integrations over only the two energy variables
 $k_0$ and $p_0$ which are the energy components of the
 four-momentum assignments $k$ and $p$, respectively, in Figs. 2
  and 3.

 According to the mechanism by means of which collinear and infrared
 divergences cancel \ref{cd},
 it is appropriate to transform from the variables $(k_0,p_0)$ to the variables
 $(K,p_0)$ where
$$
K=k_0+{\nu -p_0\over 2}
\eqno(A2)
$$
\medskip\noindent
{\bf Appendix B : Final expressions for four-particle processes}
\medskip
The convention concerning the sign of the energy of a particle
 discussed in Section~2, enables one
to view all contributing four-particle processes simultaneously
 in the two-dimensional energy plane shown in Fig. 4.
The Latin letters in Fig. 4 serve to label the subregions over which one
integrates in the $K$, $p_0$ plane. The numbers in Fig. 4
label the classes of subregions. The final analytical expressions
are different for each class.
The classes arise as follows:
For some regions of the $k_0$, $k_0^{\prime}$ plane
 ($k_0^{\prime}=k_0+q_0-p_0$), the restrictions imposed by the
 energy--momentum conserving Dirac delta function in the phase space
 integration, causes one to replace the usual
 upper limit of 1 and/or the usual lower limit of $-1$ for $\cos\theta$
 and/or $\cos\theta^{\prime}$ (where $\theta$ is the angle between $\vec{q}$
 and $\vec{k}$ and $\theta^{\prime}$ the angle between $\vec{q}$
 and $\vec{k}^{\prime}$) by expressions in terms of
 $k_0$ and $k_0^{\prime}$.
 The classes of subregions in the $k_0,\ k_0^{\prime}$ plane are distinguished
according to the
 combination of upper and lower limits
 for $z$ and $z^{\prime}$ which are obtained for each class.
 In consequence, the integrations over $z$ and $z^{\prime}$ will lead to
 different analytical results for each class.

The final results for four-particle processes {\rm after}
 the cancellation of collinear singularities are
$$
\eqalignno{
\int \bar{M}_{\Sigma }S_4d\mu
=&{1\over 4\pi ^3q_z}\int dKdp_0S_4
\left\{\left|{q^2\over 2p_0}\right|\ln|\bar{A}_4|+|q^2|\left[
{|p_0|\over 4A^2}
-{\epsilon (p_0)\over 2A}\right]\ln|\bar{B}_4|\right.  &\cr
  &+|q^2|\left[{|p_0|\over 4b^2}+{\epsilon (p_0)\over 2b}\right]
\ln|\bar{C}_4|&\cr
&+\epsilon (k_0k_0^{\prime})\left[ |E_4|+
{(B-p_0)(c-E)(C-e)\over 2A}|F_4|
\right.  &\cr
  &\left.\left.+{(a+p_0)(d+e)(D+E)\over 2b}|G_4|\right]\right\}
&(B1)\cr
}
$$
and
$$
\eqalignno{
\int \bar{M}_0S_4d\mu
=&{1\over 4\pi ^3q_z}\int dKdp_0S_4\left\{{|q^2|\over 2A}
\left[{|p_0|q^2\over 4q_z^2A}-\epsilon (p_0)
{q^2-2B\nu \over 2q_z^2}\right]
\ln|\bar{B}_4|\right.  &\cr
 & +\left|{q^2\over 2p_0}\right|{Cc+Dd\over q_z^2}\ln|\bar{A}_4|
+{|q^2|\over 2b}\left[{|p_0|q^2\over 4bq_z^2}+\epsilon (p_0)
{q^2+2a\nu\over 2q_z^2}\right]\ln|\bar{C}_4|  &\cr
 & +\epsilon (k_0k_0^{\prime}){q^2\over 2q_z^2}\left[ |E_4|
+{(B-p_0)(c-E)(C-e)\over 2A}|F_4| \right.  &\cr
 & \left.\left.+{(a+p_0)(d+e)(D+E)\over 2b}|G_4|\right]\right\},
&(B2)\cr
}$$
with quantities as defined in Table~1 and
$$
\eqalignno{
S_4=&
\left[ {1\over 2}-{\epsilon (k_0)\over 2}
\left(1-2n_F(x_k)\right)\right]
\left[ {1\over 2}+{\epsilon (k'_0)\over 2}\left(
1-2n_F(x^{\prime}_k)\right)\right]  &\cr
& \left[{1+\epsilon (p_0)\over 2}+n_B(|p_0|)\right].
&(B3)\cr
}
$$
These expressions are valid for all the four-particle processes with
the $K,p_0$ integration region given in Fig. 4.

The way in which these expressions contribute to the structure
 functions of the nucleon can be seen by considering eqs. (A2), (B1) and
 (15)--(18).
\medskip\noindent
{\bf Appendix C : Final expressions for three-particle processes}
\medskip
Similar to our discussions for the four-particle processes,
both the quark and antiquark three-particle processes can be viewed
 simultaneously in the two-dimensional energy plane shown in Fig. 5.
{}From the figure it can be seen that the $K,p_0$ integration region for
 three-particle processes is not subdivided into subregions as in the case
 of four-particle processes (see Fig. 4).
 For three-particle processes, subdivisions of the energy plane
 need only to be considered when performing angular integrations in
 intermediate steps and these subdivisions disappear after the
 angular integrations.

 The final results for three-particle processes after
 the cancellation of collinear singularities are [Declaration of
symbols appears after eq.~(C2)]:
$$
\eqalignno
{
&\sum_{W=B,F,F'} \int \bar{M}_{\Sigma}^{3 W}S_{3 W}d\mu_W          &\cr
=&{1\over 4\pi^3q_z}\int dKdp_0\theta (1-y)\theta (1+y)
\left\{S_{3 B}(-|p_0|)\left[ 1+{a\over A}+{A\over a}
+{q^2\over 2}\left({1\over a^2}+{1\over A^2}\right)\right]\right.  &\cr
 & +S_{3 F}\left[-{\epsilon (b)|q^2|\over 2p_0}
\ln |\bar{A}_F|-\epsilon (b)|q^2|\left[{p_0\over 4a^2}
-{1\over 2a}\right]\ln|\bar{C}_{\rm{ F}}|\right.  &\cr
&\left. +{\epsilon (b)|q^2|\over q_z}\left[{3\over 4}+
{a+A\over 4q_z^2}(a+A-2p_0)\right]\ln|K_{\rm{ F}}|
+|b|\left[ {1\over 2}+{A^2-a^2\over q_z^2}+{q^2\over 2a^2}
+{A\over a}\right]\right] &\cr
& +S_{3F'}\left[-{\epsilon (B)|q^2|\over 2p_0}
\ln |\bar{A}_{F'}|-\epsilon (B)|q^2|\left[{p_0\over 4A^2}-{1\over 2A}
\right]\ln|\bar{B}_{F'}|\right.  &\cr
  & -{\epsilon (B)|q^2|\over q_z}\left[{3\over 4}
+{a+A\over 4q_z^2}(a+A-2p_0)\right]\ln|K_{F'}|   &\cr
 & \left.\left.+|B|\left[ {1\over 2}
+{A^2-a^2\over q_z^2}+{q^2\over 2a^2}+{a\over A}\right]\right]
+S_2{p_0|p_0|(K+p_0/2)\over p_0^2+T^2}\right\}.
&(C1)\cr
}$$

The corresponding final result for the longitudinal polarization of the virtual
 photon is:
$$
\eqalignno
{
&\sum_{\rm{ W=B,F,F'}}\int \bar{M}_0^{3 W}S_{3 W}d\mu_W &\cr
=&{1\over 4\pi ^3q_z}\int dKdp_0\theta (1-y)\theta (1+y)&\cr
&\left\{-S_{3\rm{ B}}{|p_0|\over q_z^2}\left(-\nu^2+{q^2\over 2}\left[
 {a\over A}+{A\over a}+{q^2\over 2}\left(
{1\over a^2}+{1\over A^2}\right)
\right]\right)\right.  &\cr
 & +S_{3F}\left\{ -\epsilon (b){|q^2|\over 4p_0q_z^2}\ln |\bar{A}
_{\rm{ F}}|[q^2+4K^2+p_0^2-\nu^2]-\epsilon (b)
{|q^2|\over 2aq_z^2}
\left[ p_0\left( {q^2\over 4a}+\nu\right)\right.\right.   &\cr
 & \left.
-{q^2+2\nu a\over 2}\right]
\ln |\bar{C}_{\rm{ F}}|
 +\epsilon (b){|q^2|\over q_z}\left[
-{1\over 4}-{a+A\over 4q_z^2}(a+A-2p_0)\right]\ln|K_{\rm{ F}}| &\cr
& \left.
+\epsilon (b){b\over q_z^2}\left[ -{\nu^2\over 2}-A^2+a^2+
{q^2\over 2}\left(
{A\over a}+{q^2\over 2a^2}\right)\right]\right\}  &\cr
 & +S_{3F'}\left\{ -\epsilon (B){|q^2|\over 4p_0q_z^2}
\ln |\bar{A}_{F'}|[q^2+4K^2+p_0^2-\nu^2]-\epsilon (B)
{|q^2|\over 2Aq_z^2}\left[ p_0\left(
{q^2\over 4A}-\nu\right)\right.\right.   &\cr
 &\left.
-{q^2-2\nu A\over 2}\right]
\ln |\bar{B}_{F'}|
 -\epsilon (B)
{|q^2|\over q_z}\left[-{1\over 4}-{a+A\over 4q_z^2}
(a+A-2p_0)\right]\ln|K_{\rm{ F}^{\prime}}|  &\cr
 &\left.\left.+\epsilon (B)
{B\over q_z^2}\left[ -{\nu^2\over 2}+A^2-a^2+{q^2\over 2}\left(
{a\over A}+{q^2\over 2A^2}
\right)\right]\right\}
+S_2{p_0|p_0|(K+p_0/2)\over p_0^2+T^2}\right\}.
&(C2)\cr
}
$$
where
$$
S_{3\rm{ B}}=S_2\left[ 1+2n_{\rm{ B}}(|p_0|)\right]{1\over 2}
\eqno(C3)
$$
$$
S_{3\rm{ F}}=S_2\left[ 1-2n_{\rm{ F}}(x_{k-p})\right]{1\over 2}
\eqno(C4)
$$
$$
S_{3\rm{ F}^{\prime}}=S_2\left[ 1-2n_{\rm{ F}}(x_{k^{\prime}-p})\right]
{1\over 2}
\eqno(C5)
$$
$$
S_2=\left[{1\over 2}-{\epsilon (k_0)\over 2}\left(1-2n_{\rm{ F}}(x_k)
\right)\right]\left[{1\over 2}+{\epsilon (k_0+\nu)\over 2}\left(
1-2n_{\rm{ F}}(x_{k+q})\right)\right]
\eqno(C6)
$$
$$
x_k=|k_0|-\mu\epsilon (k_0).
\eqno(C7)
$$
$$
\bar{A}_F=-{b^2\over Dd}
\eqno(C8)
$$
$$
K_F={(q_z+\nu)(2K-p_0-q_z)\over (q_z-\nu)(2K-p_0+q_z)}
\eqno(C9)
$$
$$
\bar{C}_F={b^2\over p_0^2}
\eqno(C10)
$$
$$
\bar{A}_{F'}=-{B^2\over Dd}
\eqno(C11)
$$
$$
K_{F'}={(q_z+\nu)(2K-p_0+q_z)\over (q_z-\nu)(2K-p_0-q_z)}
\eqno(C12)
$$
$$
\bar{B}_{F'}={B^2\over p_0^2}
\eqno(C13)
$$
and where, in the step functions $\theta$,
$$
y={q^2+2k_0\nu\over 2k_0q_z} .
\eqno(C14)
$$
 The factor $S_2$ describes the statistical factors for the external
 either quarks or antiquarks in three-particle processes.
The last term in each of eqs. (C1) and (C2) has been introduced
 \ref{ezawa} to make the integrations over positive and negative $p_0$
 separately convergent for the purpose of numerical calculation. From the
 definition of $K$ it follows that $K+p_0/2$ is independent of $p_0$ and
 therefore that the last term in each of eqs. (C1) and (C2)
 contributes zero when added at a positive and negative value of $p_0$
\vfil\eject
\bigskip\centerline{\sectnfont References}\bigskip
\item{\reftag{ital})}
C. Angelini and R. Pazzi, Phys. Lett. {\bf B113} (1982) 343.
\smallskip
\item{\reftag{g5})}J. Cleymans and  R.L. Thews,
                   Z. Phys. {\bf C37} (1988) 315
\smallskip
\item{\reftag{aus})} R.P. Bickerstaff and J.T. Londergan,
                    Phys. Rev. {\bf D42} (1990) 90.
\smallskip
\item{\reftag{indians})} K. Ganesamurthy, V. Devanathan, M. Rajasekaran,
                         Z. Phys. {\bf C52} (1991) 589
\smallskip
\item{\reftag{g6})} A. Chodos et al., Phys. Rev. {\bf D9} (1974) 3471
\smallskip
\item{\reftag{zcd})} J. Cleymans and I. Dadi\'{c},
                     Z. Phys. {\bf C42} (1989) 133
\smallskip
\item{\reftag{jcd})} J. Cleymans and I. Dadi\'{c},
                     Z. Phys. {\bf C45} (1989) 57
\smallskip
\item{\reftag{ezawa})} J. Cleymans and I. Dadi\'c,
 Proceedings of the Second Workshop in Thermal Field Theories and
 Applications, Eds.  H. Ezawa et al.,
 (North Holland, Amsterdam, 1991).
\smallskip
\item{\reftag{cd})} J. Cleymans and I. Dadi\'c,
                    Phys. Rev. {\bf D47} (1993) 160.
\smallskip
\item{\reftag{bps})} R. Baier,B. Pire and D. Schiff,
		     Phys. Rev. {bf D38} (1988) 2814.
\smallskip
\item{\reftag{aab})} T. Altherr,P. Aurenche and T.Becherrawy,
		     Nucl. Phys. {bf B315} (1989) 436.
\smallskip
\item{\reftag{ab})} T. Altherr and T.Becherrawy,
		     Nucl. Phys. {bf B330} (1990) 174.
\smallskip
\item{\reftag{aa})} T. Altherr and P. Aurenche,
		     Z. Phys. {bf C45} (1990) 99.
\smallskip
\item{\reftag{glm})} T. Grandou, M. Le Bellac and J.-L. Meunier,
		     Z. Phys. {bf C43} (1989) 575.
\smallskip
\item{\reftag{ggp})} Y. Gabellini, T. Grandou and D. Poizat,
		     Ann. Phys. (N.Y.){bf 202} (1990) 436.
\smallskip
\item{\reftag{qbw})}
M. Virchaux and A. Milsztajn, Phys. Lett. {\bf B274} (1992) 221.
\smallskip
\item{\reftag{we})} J. Kwiecinski et al., Phys. Rev. {\bf D42} (1990) 3645
\smallskip
\item{\reftag{wh})} L.W. Whitlow, Ph.D. Thesis, SLAC-Report-357, March 1990
\smallskip
\item{\reftag{qbl})} J.J. Aubert et al., Phys. Lett. {\bf 105B} (1981) 315
\smallskip
\item{\reftag{g7})} R.P. Feynman, Photon-Hadron Interactions
                    (Benjamin, New York, 1972) ;
J.D. Bjorken, and E.A. Paschos, Phys. Rev. {\bf 185} (1969) 1975
\smallskip
\item{\reftag{halmar})}
F. Halzen, A.D. Martin, Quarks and Leptons (John Wiley and Sons,
New York, 1984)
\vfil\eject
\medskip
Table 1 :definitions of the quantities appearing in
 eqs. (B1) and (B2) \ref{cd}

\medskip
{\offinterlineskip \tabskip=0pt
\halign{\strut
        \vrule#&           
{}~   #  ~ &         
        \vrule#&           
        ~
        \hfil # ~ &    
        \vrule#&           
        ~
        \hfil # ~ &    
        \vrule#&           
        ~
        \hfil # ~ &    
        \vrule#&           
        ~
        \hfil # ~ &    
        \vrule#&           
        ~
        \hfil # ~ &    
        \vrule#&           
        ~
        \hfil # ~ &    
        \vrule#            
\cr
\noalign{\hrule}
&   &&  &&  &&  && && &&  &\cr
&Class && $\bar{A}_4$ && $\bar{B}_4$ && $\bar{C}_4$ && $E_4$
&& $F_4$ && $G_4$ &\cr
&   &&  &&  &&  && && &&  &\cr
\noalign{\hrule}
&1 && 1 && 1 && 1 && $p_0$&&${\displaystyle {p_0\over A(B-p_0)}}$ &&
 ${\displaystyle {p_0\over b(a+p_0)}}$ &\cr
&   &&  &&  &&  && && &&  &\cr
&2  && ${\displaystyle{B^2\over Dd}}$ &&${\displaystyle {B^2\over p_0^2}}$ &&
 ${\displaystyle {q^2\over 4dD}}$ && B &&${\displaystyle {B\over A(B-p_0)}}$&&
 ${\displaystyle {B\over (d+e)(D+E)}}$  &\cr
&   &&  &&  &&  && && &&  &\cr
&3  && ${\displaystyle{aB\over cD}}$ &&${\displaystyle {EB\over p_0c}}$ &&
  ${\displaystyle {Ea\over p_0D}}$ && E &&${\displaystyle {E\over A(c-E)}}$&&
 ${\displaystyle {E\over b(D+E)}}$ &\cr
&   &&  &&  &&  && && &&  &\cr
&4  && ${\displaystyle{cB\over ad}}$ &&${\displaystyle {Bc\over p_0E}}$ &&
 ${\displaystyle {p_0e\over ad}}$ && c &&${\displaystyle {c\over A(c-E)}}$&&
 ${\displaystyle {c\over (d+e)(a+p_0)}}$  &\cr
&   &&  &&  &&  && && &&  &\cr
&5  && ${\displaystyle{aD\over BC}}$ &&${\displaystyle {p_0-e}\over C}$ &&
 ${\displaystyle {aD\over p_0E}}$ && D &&${\displaystyle{D\over
(B-p_0)(C-e)}}$&&
 ${\displaystyle {D\over b(D+E)}}$  &\cr
&   &&  &&  &&  && && &&  &\cr
&6  && ${\displaystyle{aB\over Cd}}$ &&${\displaystyle {eB\over p_0C}}$ &&
 ${\displaystyle {ea\over p_0d}}$ && $p_0-e$ &&${\displaystyle {p_0-e\over
(C-e)(B-p_0)}}$
&&${\displaystyle {p_0-e\over (a+p_0)(d+e)}}$  &\cr
&   &&  &&  &&  && && &&  &\cr
&7  && ${\displaystyle{a^2\over Cc}}$ &&${\displaystyle {q^2\over 4Cc}}$ &&
${\displaystyle {a^2\over p_0^2}}$ && a &&${\displaystyle {a\over (C-e)(c-E)}}$
 &&${\displaystyle {a\over b(a+p_0)}}$  &\cr
&   &&  &&  &&  && && &&  &\cr
\noalign{\hrule}
}
}
\vfil\eject
\bigskip\centerline{\sectnfont Figure Captions}\bigskip
\item{Fig. 1}
Processes contributing to deep inelastic scattering
to first order in $\alpha_s$ .
\smallskip
\item{Fig. 2}
Diagrams for four-particle processes.
\smallskip
\item{Fig. 3}
Diagrams for three-particle processes.
\smallskip
\item{Fig. 4}
Regions of support for all four-particle processes in the $K$, $p_0$ plane
 \ref{cd}.
\smallskip
\item{Fig. 5}
Regions of support for all three-particle processes in the $K$,$p_0$
 plane
\smallskip
\item{Fig. 6}
Plots of $F_2$ versus $x$ at $Q^2=4$ GeV$^2$; solid
line reproduces the
 parton distributions of Kwiecinski et al. \ref{we} ($B_-$ fit)
 and dashed line represents the zeroth order in $\alpha_s$
 expression given in eq. (10).
\smallskip
\item{Fig. 7}
Plot of $R$ versus $x$ at $Q^2=4$ GeV$^2$ for
$\alpha_s=0.2$, $T=0.04$  GeV,
$\mu_{up}=0.21$ GeV and
$\mu_{down}=\mu_{up}/2^{1\over 3}$
\smallskip
\item{Fig. 8}
Plots of $F_2$ versus $x$ at $Q^2=4$  GeV$^2$;
short dashed line represents the zeroth order in $\alpha_s$
 expression given in eq. (10) and
the long dashed line represents the zeroth
 order plus ${\cal O}(\alpha_s$ expression given in eqs.
 (8) and (15) with
 $\alpha_s=0.2$.
The solid line is the $B_-$ fit of Ref. \ref{we}.
\smallskip
\item{Fig. 9}
Plots of $F_2$ versus $x$ at $Q^2=4$  GeV$^2$ as a function of temperature
 at fixed chemical potential $\mu_{up}=0.133$ GeV
according to the zeroth
 order plus ${\cal O}(\alpha_s)$ expression given in eqs.
 (8) and (15) with
 $\alpha_s=0.2$. The temperature is 0.043 GeV for the lowest dashed curve
0.055 GeV for the next upper one, 0.067 GeV for the third dashed line and 0.079
GeV for the highest dashed line.
The solid line is the $B_-$ fit of Ref. \ref{we}.
\smallskip

\item{Fig. 10}
Plots of $F_2$ versus $x$ at $Q^2=4$  GeV$^2$ as a function of temperature
 at fixed temperature $T=0.067$ GeV
according to the zeroth
 order plus ${\cal O}(\alpha_s)$ expression given in eqs.
 (8) and (15) with
 $\alpha_s=0.2$. The
chemical potential
$\mu_{up}$ is 0.109 GeV for the lowest dashed curve
0.133 GeV for the next upper one,
0.157 GeV for the third dashed line and 0.181 GeV
for the highest dashed line.
The solid line is the $B_-$ fit of Ref. \ref{we}.
\smallskip
\item{Fig. 11}
Plot of $R$ versus $x$ at $Q^2=4$ GeV$^2$ with
$\alpha_s=0.2$, $T=0.067$ GeV,
$\mu_{up}=0.133$ GeV and
$\mu_{down}=\mu_{up}/2^{1\over 3}$
\smallskip
\vfil\eject\bye